\documentclass[paper,notoc]{JHEP3}

\JHEPspecialurl{http://jhep.sissa.it/JOURNAL/JHEP3.tar.gz}
\usepackage{graphics}

 \newcommand{\be}{\begin{eqnarray}}
 \newcommand{\ee}{\end{eqnarray}}
 \newcommand{\nee}{\nonumber\end{eqnarray}}
 \newcommand{\nn}{\nonumber\\}
 
 \newcommand{\plr}{\stackrel{\leftrightarrow}{\partial}{}\!\!}

\title{Supersymmetry through CP violation in $H^\pm \to W^\pm\,h^0$}

\author{E. Christova, E. Ginina, M. Stoilov\\
    Institute of Nuclear Research and Nuclear Energy,
        Sofia 1784, Bulgaria\\
    E-mail: \email{echristo@inrne.bas.bg, eginina@inrne.bas.bg, mstoilov@inrne.bas.bg}}

\abstract{The CP violating asymmetry between the partial decay
rates of $H^+ \to W^+ h^0$ and $H^- \to W^- h^0$ is calculated in
first order of the weak  coupling constant $\alpha_{\omega}=g^2/4
\pi$ in MSSM with complex parameters. The dependence on the phases
of $A_\tau$ and $M_1$ is discussed. Different values of $\tan
\beta$ are considered. The asymmetry is up to the order of
$10^{-2} $.}

\keywords{Supersymmetric Standard Model, Higgs Physics, CP violation}

\begin{document}

\section{Introduction}

    The Minimal Supersymmetric Standard Model (MSSM) implies the
existence of a pair of charged Higgs bosons $H^\pm$. At tree
level there are three possible decay modes of $H^\pm$  into
ordinary particles: $H^+ \to t\bar b$, $H^+ \to \nu\tau^+$ and
$H^+ \to W^+h^0$ where $h^0$ is the lightest neutral Higgs boson.
The lighter fermions from the first two generations and the
heavier neutral Higgs bosons  are not considered.   Loop
corrections due to a Lagrangian with CP violating phases lead to
decay rate asymmetries between the partial decay widths of $H^+$
and $H^-$ and that would be a clear signal of CP violation. In
refs.\cite{1} and \cite{2} such decay rate asymmetries are
considered in the MSSM with complex phases for the quark decay
mode $H^+ \to tb$ -- the asymmetry $\delta_{tb}$, and for the
lepton decay mode $H^+ \to \nu\tau$ -- the asymmetry
$\delta_{\nu\tau}$.
 In order to finalize this investigation we consider
here the decay rate asymmetry of $H^\pm \to W^\pm h^0$:
\be \delta_{Wh^0} =\frac{\Gamma (H^+ \to W^+h^0) - \Gamma (H^- \to
W^-h^0)} {\Gamma (H^+ \to W^+h^0) + \Gamma (H^- \to
W^-h^0)}.\label{111}
\ee

We shall work in MSSM with complex parameters in first order
 of the weak coupling constant
$\alpha_{\omega}=g^2/4 \pi$. Within MSSM, after redefining the
fields,
 the new sources of CP violation are the
phase of the higgsino mass parameter $\mu$,  two of the phases of the
gaugino masses $M_{i}$, $i = 1,2,3$ (we choose these to be the phases of $M_1$ and $M_3$),
 and the phases of the trilinear
couplings of the fermions $f$, $A_{f}$. Especially the latter ones
are practically unconstrained. Previously this asymmetry was
considered in the two-Higgs doublet model in \cite{Lavoura}.

 Discussing CP violation in the decay widths, we
must keep in mind the branching ratios of the relevant decay
modes. $H^+\to \nu\tau^+$ is significant for low $m_{H^+}$, below
the $\bar tb$ threshold. This determines the sensitivity of
$\delta_{\nu\tau}$ to light $\tilde\nu \tilde\tau$ and
$\tilde\chi^0 \tilde\chi^\pm$ in the loops, i.e. to the phases
$\phi_\tau$ and $\phi_1$ of $A_\tau$ and $M_1$. The decay $H^+\to
t\bar b$ dominates for high $m_{H^+}$, which determines the
sensitivity of $\delta_{tb}$ to the phases of $A_t$ and $A_b$.
The complication with the decay $H^+\to W^+h^0$ is that the final
state $h^0$ is not observed yet and $m_{h^0}$ is an unknown
parameter. However, once $m_{H^+}$ and $\tan\beta$ are fixed, the
SUSY structure of the theory determines uniquely both $m_{h^0}$
and the coupling $H^+W^-h^0$. There are two consequences that are
important for us. First, increasing $m_{H^+}$, the mass
 $m_{h^0}$ is saturated approaching its maximum value.
 At tree level this is particularly simple:
 $m_{h^0}\leq  m_{h^0}^{max} = m_Z\vert\cos 2\beta\vert$
 \cite{Gunion&Haber}, while including QCD and SUSY
 radiative corrections $ m_{h^0}^{max}$ can be increased
 considerably,  $m_{h^0}^{max}\simeq$
 130 GeV (for a recent review see \cite{review} and the refs.
 therein).
 There is also an experimental lower bound for $m_{h^0}$, $m_{h^0}\geq 96 $ GeV
 \cite{LEP-h0}.
 Thus, respecting both the experimental and theoretical bounds,
  we shall consider $m_{h^0}$ in
the range 96 $ \leq m_{h^0}\leq$ 130 GeV. In this range of $m_{h^0}$,
for
 $m_{H^+} >m_W + m_{h^0}$ we are already in the
saturation regime where we may keep $m_{H^+}$ and $m_{h^0}$ as
independent parameters. The second consequence concerns the
$H^+W^-h^0$ coupling which determines the Br($H^+\to W^+h^0)$.
Increasing $m_{H^+}$ it quickly falls down and, depending on
$\tan\beta$,
 we can enter the so called decoupling limit, $\cos^2 (\beta -\alpha)\to 0$,
 where the Br($H^+\to W^+h^0)$ almost vanishes. This imposes severe restrictions on
 $m_{H^+}$ and $\tan\beta$. In order to keep the value of Br($H^+\to W^+h^0)$ at the level of few percents,
 we shall consider 200 $\leq m_{H^+}\leq 600$ GeV and low $\tan\beta$,
 $3 \leq \tan\beta \leq 9$ ($\tan\beta \leq 3$ is already excluded from
 the Higgs searches at LEP\cite{LEP-h0}). In accordance with this,
 $\delta_{Wh^0}$ will receive contributions from
$\tilde\nu \tilde\tau^\pm$, $\tilde t\tilde b$ and $\tilde\chi^0
\tilde\chi^\pm$ in the loops. Due to the large mass of the
top-quark, the radiative corrections with stops and sbottoms with
low masses appear to be too large to be considered within the
$\alpha_w$ approximation used here. This means that we assume that the squarks are
heavy and will not contribute in the considered range of
$m_{H^+}$. This will be discussed in the
next Section 2. Thus, we shall consider the
sensitivity of the asymmetry $\delta_{Wh}$ to the phases
$\phi_\tau$ and $\phi_1$ of the chargino-neutralino and the
slepton sectors. According to the experimental limits on the
electric dipole moments of the electron and the neutron, we
assume  a zero phase for the Higgsino mass parameter $\mu$,
$\phi_\mu =0$. However, as is shown in \cite{phasemu}, a
large phase $\phi_\mu$ is not impossible, it would require
 fine-tuning between the phase
$\phi_\mu$ and the other SUSY parameters. We end up with a short
discussion of the influence of $\phi_\mu$ on $\delta_{Wh^0}$.

 The paper is
organized as follows. In the next section we present the analytic
expression for the
 asymmetry. In Section 3 we discuss the numerical results in MSSM.
 We end up with a conclusion. There are two Appendicies - with
 the Lagrangian and with the
 analytic expressions for the
imaginary parts of the Passarino Veltmann (PV) integrals.

\section{The asymmetry}

We write the matrix elements of $H^+\to W^+h^0$ and $H^-\to
W^-h^0$ in the form:
 \be
M_{H^\pm}=ig\varepsilon_{\alpha}^{\lambda}(p_W)p_h^{\alpha}Y^\pm,
\ee%
where $\varepsilon_{\alpha}^{\lambda}(p_W)$ is the
polarization vector of $W^\pm$, $Y^\pm$ are the loop corrected
couplings:
 \be
 Y^\pm=y+\delta Y_1^\pm+ \delta Y_2^\pm+ \delta
Y_3^\pm+....
\ee%
Here $y$ is the tree level coupling:
 \be
 y=\cos
(\alpha - \beta),
\ee
and $\delta Y_k^\pm, k=1,2,3,...$ are the SUSY-induced loop
corrections.\\
The decay rates of $H^\pm\longrightarrow W^\pm h^0$ are:
 \be
\Gamma (H^\pm \longrightarrow W^\pm h^0)= {\alpha_{\omega}\over
16}{\lambda^{3/2}(m_H^2, m_h^2, m_W^2)\over m_H^3m_W^2}|Y^\pm |^2
\label{222}
\ee
Here:
 \be
\lambda (x, y, z)&=&x^2+y^2+z^2-2xy-2xz-2yz,\\
|Y^\pm\vert^2 &=&y^{2}+2y \sum _k\Re e (\delta Y_k ^\pm)+{\cal
O}(\alpha_{\omega}^2). \label{333}
\ee
At tree level the decay
widths of $H^+$ and $H^-$ are equal and there is no CP violation:
\be
\Gamma_{tree}(H^\pm\longrightarrow W^\pm h^0)=
{\alpha_{\omega}\over 16}{\lambda^{3/2} (m_H^2, m_h^2,
m_W^2)\over m_H^3m_W^2}\cos^2(\alpha-\beta).
 \ee
 CP violation is
induced by loop corrections.  They  have
CP-invariant and CP-violating contributions:
\be
\delta
Y_k^{\pm}=\delta Y_k^{inv} \pm \delta Y_k^{CP}
\ee
 and each
one has real and imaginary parts:
 \be
\delta Y_k^{inv}=\Re e (\delta Y_k^{inv})+i\Im m (\delta
Y_k^{inv}),  \quad  \delta Y_k^{CP}=\Re e (\delta Y_k^{CP})+i\Im m
(\delta Y_k^{CP})
\ee
The asymmetry $\delta_{Wh}$ is determined by $\Re e (\delta
Y_k^{CP})$. Further we shall work in first order of the weak
radiative coupling constant $\alpha_{\omega}=g^2/4 \pi$. This
approximation means that we neglect the CP-invariant loop
corrections $\Re e (\delta Y_k^{inv})$ in the denominator in
formula ($\ref{111}$). Then from ($\ref{111}$), ($\ref{222}$) and
($\ref{333}$) we obtain:
\be
 \delta_{Wh^0}\simeq
 {2\sum \limits_k \Re e(\delta Y_k^{CP})\over y }.
\ee
Here the sum is over the loops with CP-violation. We work in MSSM
with complex parameters. The Lagrangian is in Appendix A.

 In
general, there are two types of SUSY radiative corrections --
self-energy loops and  vertex corrections. We have self-energy
loops on the $W^+$-line, on the $h^0$-line and on the $H^+$-line.
The loops on the $W^+$-line are proportional to the 4-momentum of
$W^+$, $p_W^\alpha$, and because of the gauge invariance
($p_W^\alpha\epsilon_\alpha (p_W) =0$),  equal to zero. As
$\delta_{Wh^0}$
 is determined by the absoptive parts of the loops, the
loops on the $h^0$-line  will contribute  only
if the kinematic condition $m_{h^0}^2\geq (\widetilde {m}_1 +
\widetilde {m}_2)^2$ is satisfied, where $\widetilde {m}_1$
 and $\widetilde {m}_2$ are the  masses of the two particles in the loop.
However, because of the upper theoretical bound on $m_{h^0}$
($m_{h^0}\leq 130 GeV$) and the lower experimental bounds on the
SUSY particles, the above condition cannot be fulfilled for any
pair of SUSY particles $(\widetilde {m}_1 + \widetilde {m}_2)$
and these loops will not contribute either. Thus, the only
radiative corrections that will contribute are the self-energy
loops on the $H^+$-line and all vertex corrections.

 According to
the particles in the loops, we have radiative corrections with
sneutrinos and staus (Fig.1a with sleptons), with stops and
sbottoms (Fig.1a with squarks) and with charginos and neutralinos
(Fig.1b). (Note that the radiative corrections with ordinary
quarks and with Higgs bosons are CP invariant.) The radiative
corrections with scalar quarks are proportional to $m_t$ and, in
general, one would expect that they should give the main
contribution to $\delta_{Wh^0}$. This contribution is enhanced,
in addition, by the colour factor 3 that multiplies each diagram
with squarks. Having at our disposal the parameters of the scalar
mass matrices, one can achieve masses of the squarks that are low
enough to be kinematically allowed in the considered range of
$m_{H^+}$,  $m_{H^+}^2\geq ({ m}_{\tilde b_m} +
 m_{\tilde t_n})^2$, still respecting the experimental
bounds. However, at such small masses of the squarks,  the
CP-invariant radiative corrections to the denominator in
($\ref{111}$)  also grow and one can no longer expect that our
first order formula (2,10) would be a good approximation.
 The performed numerical analysis  confirmed these arguments.
 That's why we shall not
consider the contribution of loops with squarks. Physically, this
means that we assume that they are heavy and the decay $H^+ \to
\tilde t \tilde b$ is not allowed kinematically. In the commonly
discussed models of SUSY breaking, the squarks are much heavier
than  sleptons,  charginos and neutralinos.

 Thus, finally we are left with the loop corrections on Fig.1a with sleptons and
 on Fig. 1b with chargino and neutralinos. The full analytic expressions
for $\delta Y_k$ are rather lengthy but they considerably
symplify for $\Re e (\delta Y_k^{CP})$.

For the loops with staus and sneutrinos we have:
 \be
 \Re e(\delta
Y_1^{CP})(\widetilde{\tau}\widetilde{\nu}\widetilde{\nu})&=&
\frac{\alpha_{\omega}m_z\sin(\alpha+\beta)}{ 8\pi m_W  \cos
\theta_W}\sum_{m=1,2}\Im m((g_4^{\widetilde{\tau}})_m {\cal
R}_{Lm}^{\widetilde{\tau} *})\Im
m(C_0^{(1)}+C_1^{(1)}+C_2^{(1)})\nn C_X^{(1)}&=&C_X(m_H^2, m_W^2,
m_h^2, m_{\widetilde{\nu}}^2, m_{\widetilde{\tau}_m}^2,
m_{\widetilde{\nu}}^2) \ee
\be \Re e(\delta
Y_2^{CP})(\widetilde{\nu}\widetilde{\tau}\widetilde{\tau})&=&
-{\alpha_{\omega }\ \over 4\pi m_W}\sum_{m,n=1,2}\Im
m((g_4^{\widetilde{\tau}})_n {\cal
R}_{Lm}^{\widetilde{\tau}*}c_{mn}^{\widetilde{\tau}})\Im m
(C_0^{(2)}+C_1^{(2)}+C_2^{(2)})\nn C_X^{(2)}&=&C_X(m_{H^+}^2,
m_W^2, m_h^2, m_{\widetilde{\tau}_n}^2, m_{\widetilde{\nu}}^2,
m_{\widetilde{\tau}_m}^2) \ee
\be
 \Re e(\delta
Y_3^{CP})(\widetilde{\tau}\widetilde{\nu})
&=&{\alpha_{\omega}\over 8\pi m_{H^+}^2
m_W^2}\sin(\beta-\alpha)\sum_{m=1,2}
 (m^2_{\tilde \tau_m}-m^2_{\tilde\nu})\Im
m((g_4^{\widetilde{\tau}})_m{\cal R}_{Lm}^{\widetilde{\tau}*})\Im
m \,B_0^{(3)}\nn B_0^{(3)}&=&B_0(m_{H^+}^2,
m_{\widetilde{\nu}}^2, m_{\widetilde {\tau}_m}^2)
 \ee

The loops with charginos and neutralinos give:
\be \Re e(\delta
Y_4^{CP})(\widetilde{\chi}^+\widetilde{\chi}^0\widetilde{\chi}^0)
 & =&-{\alpha_{\omega}\over 2\pi }\sum_{{i=1,2}\atop {k,l=1,2,3,4}}\left\{ \Im
m(f_{ik}^LA_{kl}O_{li}^R+f_{ik}^RA_{kl}^*O_{li}^L)\right.\left[m_{\widetilde{\chi}_k^0}^2\Im
m(2C^{(4)}_0+C_1^{(4)}+C_2^{(4)})\right.\nn &&  \qquad \qquad
\qquad \qquad \qquad \left.+\,m_{H^+}^2\Im m(C_1^{(4)})+m_h^2\Im
m(C_2^{(4)})\right]\nn && \quad +\,m_{\widetilde{\chi}_k^0}
m_{\widetilde{\chi}_i^+}\Im
m(f_{ik}^LA_{kl}^*O_{li}^L+f_{ik}^RA_{kl}O_{li}^R)\Im
m(C^{(4)}_0+C_1^{(4)}+C_2^{(4)})\nn && \quad
+\,m_{\widetilde{\chi}_k^0} m_{\widetilde{\chi}_l^0}\Im
m(f_{ik}^LA_{kl}^*O_{li}^R+f_{ik}^RA_{kl}O_{li}^L)\Im
m(C^{(4)}_0+C_1^{(4)}+C_2^{(4)})\nn && \quad
\left.+\,m_{\widetilde{\chi}_i^+} m_{\widetilde{\chi}_l^0}\Im
m(f_{ik}^LA_{kl}O_{li}^L+f_{ik}^RA_{kl}^*O_{li}^R)\Im
m(C_1^{(4)}+C_2^{(4)})\right\}\nn C_{X}^{(4)}&=&C_X(m_{H^+}^2,
m_W^2, m_h^2, m_{\widetilde{\chi}_k^0}^2,
m_{\widetilde{\chi}_i^+}^2, m_{\widetilde{\chi}_l^0}^2). \ee
\be \Re e(\delta
Y_5^{CP})(\widetilde{\chi}^0\widetilde{\chi}^+\widetilde{\chi}^+)
&= &{\alpha_{\omega}\over 2\pi }\sum_{{i,j=1,2}\atop
{k=1,2,3,4}}\left\{ \Im
m(f_{ik}^LO_{kj}^L\widetilde{A}_{ij}+f_{ik}^RO_{kj}^R\widetilde{A}_{ji}^*)[m_{\widetilde{\chi}_i^+}^2\Im
m(2C^{(5)}_0+C_1^{(5)}+C_2^{(5)})\right.\nn &&
\qquad\qquad\qquad\qquad\qquad  +\,m_{H^+}^2\Im
m(C_1^{(5)})+\,m_h^2\Im m(C_2^{(5)})]\nn && \quad
+\,m_{\widetilde{\chi}_i^+} m_{\widetilde{\chi}_j^+}\Im
m(f_{ik}^LO_{kj}^L\widetilde{A}_{ji}^*+f_{ik}^RO_{kj}^R\widetilde{A}_{ij})\Im
m(C^{(5)}_0+C_1^{(5)}+C_2^{(5)})\nn && \quad
+\,m_{\widetilde{\chi}_k^0} m_{\widetilde{\chi}_i^+}\Im
m(f_{ik}^LO_{kj}^R\widetilde{A}_{ji}^*+f_{ik}^RO_{kj}^L\widetilde{A}_{ij})\Im
m(C^{(5)}_0+C_1^{(5)}+C_2^{(5)})\nn
 &&\quad \left.+\,m_{\widetilde{\chi}_k^0}
m_{\widetilde{\chi}_j^+}\Im
m(f_{ik}^LO_{kj}^R\widetilde{A}_{ij}+f_{ik}^RO_{kj}^L\widetilde{A}_{ji}^*)\Im
m(C_1^{(5)}+C_2^{(5)})\right\}\nn C_{X}^{(5)}&=&C_X(m_{H^+}^2,
m_W^2, m_h^2, m_{\widetilde{\chi}_i^+}^2,
m_{\widetilde{\chi}_k^0}^2, m_{\widetilde{\chi}_j^+}^2). \ee
\be \Re e(\delta Y_6^{CP})(\widetilde{\chi}^+\widetilde{\chi}^0)
& =& -{\alpha_{\omega}\over 4\pi}{\sin (\beta-\alpha)\over
m_{H^+}^2 m_W} \sum_{{i=1,2}\atop {k=1,2,3,4}}
\{m_{\widetilde{\chi}_k^0}
(m_{H^+}^2+m_{\widetilde{\chi}_i^+}^2-m_{\widetilde{\chi}_k^0}^2)\Im
m (f_{ik}^LO_{ki}^R+f_{ik}^RO_{ki}^L) \nn &&  \qquad \qquad \quad
-\,m_{\widetilde{\chi}_i^+}(m_{H^+}^2-m_{\widetilde{\chi}_i^+}^2+
m_{\widetilde{\chi}_k^0}^2)\Im
m(f_{ik}^LO_{ki}^L+f_{ik}^RO_{ki}^R)\}\Im m(B_0^{(6)})\nn
B_0^{(6)}&=&B_0(m_{H^+}^2, m_{\widetilde{\chi}_k^0}^2,
m_{\widetilde {\chi}_i^+}^2) \ee
Here the imaginary parts of the PV
integrals $C_X(m_{H^+}^2,m_W^2, m_h^2,m_0^2,m_1^2,m_2^2)$ and \\
$B_0(m_{H^+}^2,m_0^2,m_1^2)$ enter. The appropriate analytic
expressions are given in Appendix B.

\section{Numerical Results}

%
%
\begin{figure}[t]
\includegraphics{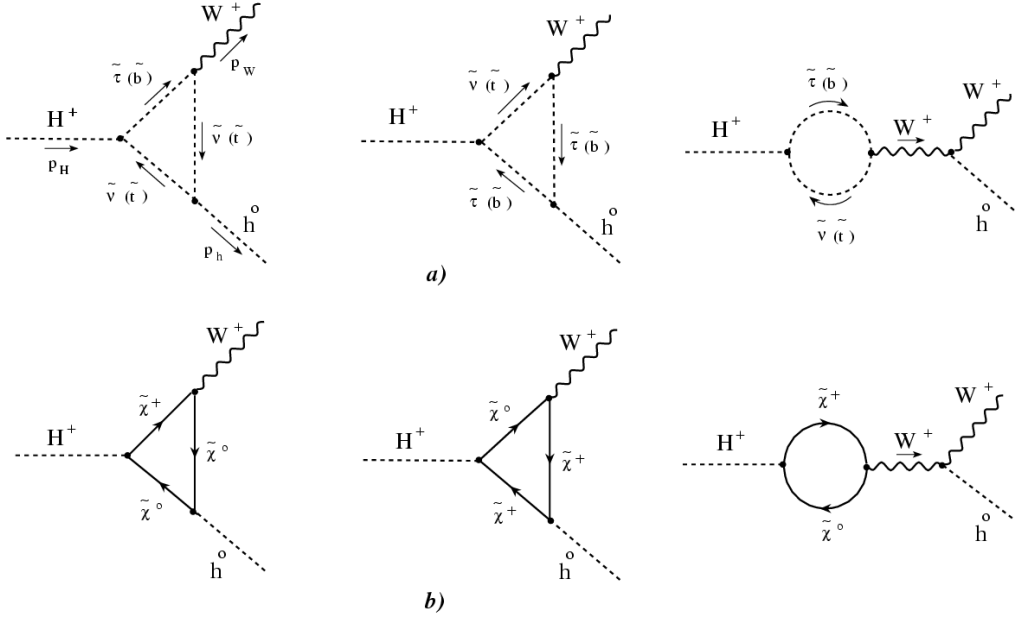} \caption{ The 1-loop diagrams in MSSM
with complex parameters that contribute to $\delta_{Wh^0}$}
\end{figure}
\begin{table}[b]\center
\begin{tabular}{|c|r|r|r|r|r|r|r|r|r|}
  \hline
$\tan \beta$ & $m_{\tilde\nu}$& $m_{\tilde\tau_1}$ &
$m_{\tilde\tau_2}$ & $m_{\tilde\chi^+_1}$
  &  $m_{\tilde\chi^+_2}$ &  $m_{\tilde\chi^0_1}$ & $m_{\tilde\chi^0_2}$
  & $m_{\tilde\chi^0_3}$& $m_{\tilde\chi^0_4}$ \\
  \hline
  $3$ & 105 & 119 & 130 & 116 & 291 & 96 & 139 & 162 & 291\\
  \hline
   $6$ & 102 & 118 & 133 & 123 & 288 & 100 & 139 & 167 & 287 \\
   \hline
   $9$ & 102 & 115 & 136 & 126 & 286 & 101 & 139 & 169 & 286\\ \hline
\end{tabular}
\caption{The masses of the superparticles for the parameters
(3.1) and $\phi_\tau=\phi_1=\pi/2$.}
 \end{table}
\begin{figure}[b]
\includegraphics{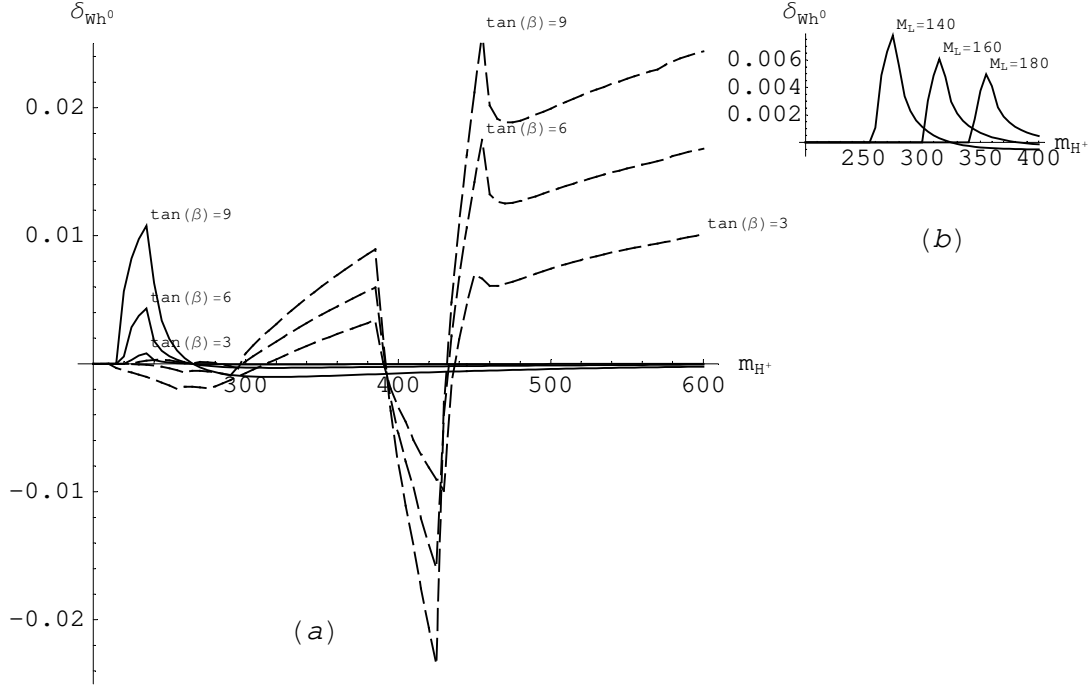}
\caption{$~\delta_{Wh^0}$ as a function of $m_{H^+}$ {\bf a)~} for
different values of $\tan\beta$, $M_L=120$~GeV; solid lines are
for $\phi_\tau=-\pi/2,~\phi_1=0$; dashed lines are for
$\phi_\tau=0,~\phi_1=-\pi/2$. {\bf b)~} for different values of
$M_L$, at $\tan\beta=9,~\phi_\tau=-\pi/2,~\phi_1=0$. }
\end{figure}
\begin{figure}[t]
\includegraphics{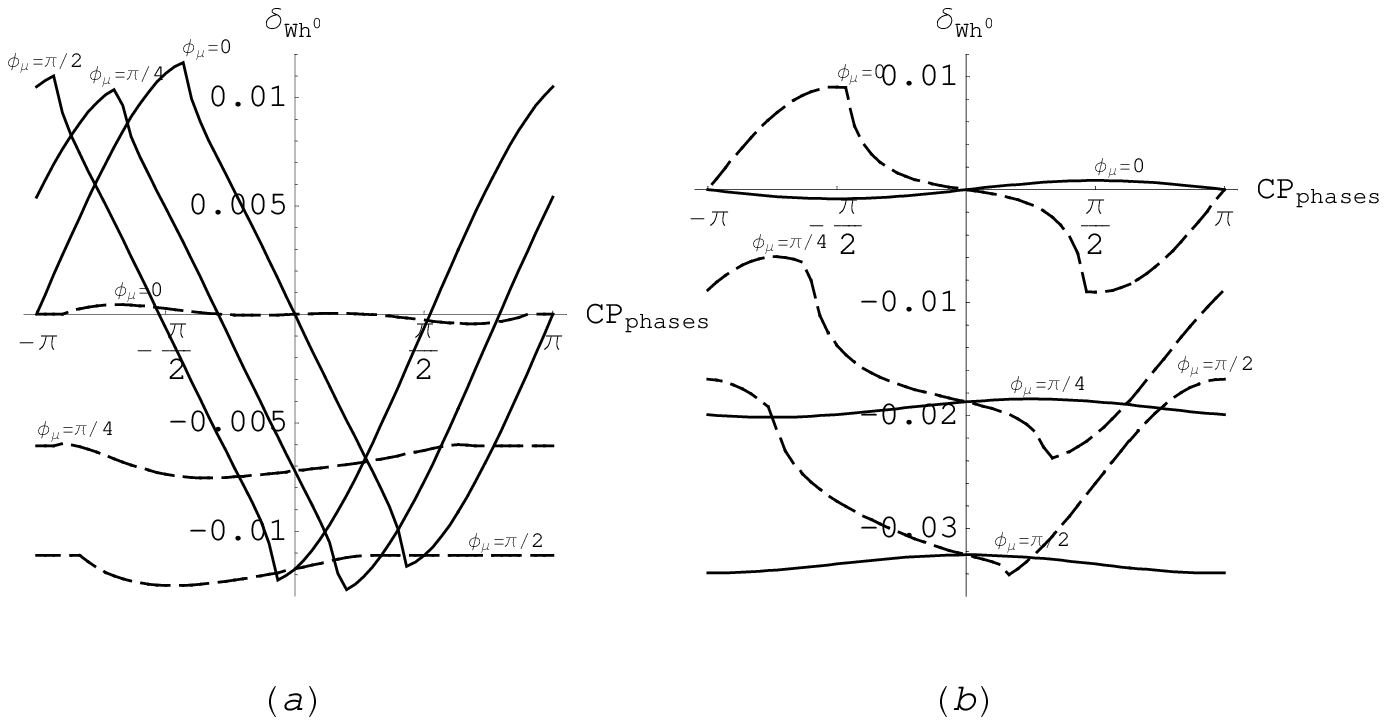}
\caption{$~\delta_{Wh^0}$ at $\tan\beta=9$ versus the CP violating
phases $\phi_\tau$ and $\phi_1$ for different values of
$\phi_\mu$ [ $\phi_\mu = 0, \pi /4, \pi /2$]. The solid lines are
for $\phi_\tau = [-\pi , \pi ]$ while $\phi_1=0$; the dashed lines
are for $\phi_1 = [-\pi , \pi ]$ while $\phi_\tau =0$,
 {\bf a)~} for $m_{H^+}=237$~GeV
($=m_{\tilde\nu}+m_{\tilde\tau^+_2})$ and
 {\bf b) ~}for
 $m_{H^+}=387$~GeV
($=m_{\tilde\chi^+_2}+m_{\tilde\chi^0_1})$. }
\end{figure}

Here we present our numerical analysis for the dependence of
$\delta_{Wh^0}$ on the MSSM parameters. Taking into account the
lower experimental  and the upper theoretical bounds on the mass
$m_{h^0}$, we consider  the range $96\leq m_{h^0}\leq 130$ GeV.
Our analysis showed a very weak dependence on $m_{h^0}$ and the
results here are presented for
$$m_{h^0}=125 ~{\rm GeV.}$$

As explained in  Section 2 the diagrams with squarks
cannot be considered within the $\alpha_w$-approximation used here
 and  our analysis will be based on the diagrams on Fig. 1a with
sleptons, and on Fig.1b. Thus our numerical results  will depend
on the parameters of the slepton, chargino and neutralino
sectors. In order not to vary too many parameters we fix part of
the SUSY parameter space:
 \be
  M_2=250~{\rm GeV},~
     M_{ E} = M_{ L} - 5~{\rm GeV},M_L=120~{\rm GeV},\nn
|A_\tau| = 500~{\rm GeV},~
     \vert \mu \vert=150~{\rm GeV.}~\label{leppars}
        \ee

Assuming  the GUT relation only for the absolute values of
$M_1$ and $M_2$
 $(\vert
M_1\vert =\frac{5}{3}\tan\theta_W\,\vert M_2\vert$), we keep
$\phi_1$, the phase of $M_1$, as a physical phase. The phase of
$M_2$ can be rotated away. The other  phase relevant for our
considerations is the phase $\phi_\tau$ of $A_\tau$. The varied
parameters thus, are the charged Higgs mass, $\tan\beta$ and the
CP-violating phases. We consider $\tan\beta$  in the interval
$$3\leq\tan\beta \leq 9.$$

As it is well known, in order $\delta_{Wh^0}$ to be nonzero we
need both new decay channels  opened and  CP violating phases.
 In accordance with
this we have three cases: i) When only the decay channels $H^+\to
\tilde\nu \tilde\tau^+_n$ are open (Fig. 1a). Then the phase
$\phi_\tau$ is responsible for CP violation. ii) When the decay
channels $H^+\to\tilde\chi^+_i \tilde\chi^0_k$ are open only
(Fig. 1b). In this case CP violation is due to the phase
$\phi_1$, and iii) When both $H^+\to \tilde\nu \tilde\tau^+_n$ and
$H^+\to\tilde\chi^+_i \tilde\chi^0_k$ decay channels are
kinematically allowed ( all diagrams of Fig.1). In this case the
two phases $\phi_\tau$ and $\phi_1$ contribute.

Examples of the asymmetry as function of $m_H^+$ for cases i) and
ii) are shown on Fig.2a for  different values of $\tan\beta$. The
relevant SUSY mass spectrum is presented in Table 1. Numerically
the two cases are obtained taking $\phi_\tau \neq 0$, $\phi_1$=0
for case i),  and
 $\phi_\tau$ = 0, $\phi_1 \neq 0$ for case ii)\footnote{The two
  cases i) and ii) can, surely, be obtained  alternatively
  varying the mass parameters $M_2,~\mu$ and $ M_L$.}.
  It is clearly seen that in both cases
 the asymmetry strongly increases with $\tan\beta$. The positions of
the spikes correspond to the threasholds of the decay modes to the
intermediate particles in the loops. It is  seen that
 for $m_{H^+} \leq $ 300 GeV the asymmetry is dominated by
the light staus and sneutrinos in the SUSY mass spectrum - the
solid lines on Fig. 2a, while for $m_{H^+} \geq $ 300 GeV  the
asymmetry is determined by the charginos and neutralinos - the
dashed lines on Fig. 2a. In both cases, i) and ii), the asymmetry
reaches up to $10^{-2}$. The dependence on  $M_L$ for case i) is
seen on Fig.2b.  Case iii), when all relevant SUSY particles can
be light, is described  by
 the algebraic sum of the two graphs at a given $\tan\beta$
 and we don't present it separately. In all these cases the asymmetry does not exceed
 few percents.

 Up to now, all presented results are for the phase of
$\mu$  zero, $\phi_\mu  =0$, i.e. for positive $\mu$'s. Negative
$\mu$ roughly speaking just flips the sign of $\delta_{Wh^0}$.
 The effect of a non zero phase $\phi_\mu$ is seen on Fig.3, where
  we show the dependence of
the asymmetry on $\phi_\tau$  (solid lines), and on
$\phi_1$  (dashed lines) for $\phi_\mu = 0,\pi/4$
and  $\pi/2$.
 The values of $m_{H^+}$ are chosen to be near the
threasholds of the  decay channels: $m_{H^+}=
m_{\tilde\nu}+m_{\tilde\tau^+_2}=237$~GeV on Fig.3a, and $m_{H^+}=
m_{\tilde\chi^+_2}+m_{\tilde\chi^0_1}=378$~GeV on Fig.3b. In all
cases a CP violating phase of $\mu$
 does not change the form of the curves but rather  shifts
 the positions of the maximum and, in general, increases the absolute value of the asymmetry.
 Note that  even for
 $\phi_\mu=0$ the maximal effect is not achieved for
$\phi_\tau ( \phi_1 ) =-\pi/2$.

\section{Conclusions}

We have considered the CP violating asymmetry $\delta_{Wh^0}$ of
the decay rate difference between $H^+\to W^+ h^0$ and $H^-\to
W^- h^0$ induced by  one loop radiative corrections  in MSSM with
complex parameters. This decay is important for relatively low
$m_{H^+}$ and $\tan\beta$. This in turn determines the importance
of $\delta_{Wh^0}$ only if there are relatively low SUSY masses.
We have considered the contribution from $m_{\tilde\nu}$ and
$m_{\tilde\tau}$, and/or $m_{\tilde\chi^+}$ and
$m_{\tilde\chi^0}$ in the loops, and thus the sensitivity to the
phases of $A_\tau$, $M_1$ and $\mu$. We work in first order of the
weak coupling constant $\alpha_w$ and this approximation is not
enough to consider the contribution of light stops and sbottoms.
 Typical values for the asymmetry are about
~ $10^{-2}\div10^{-3}$, the main contributions being from
$\tilde\nu$ and $\tilde\tau$ for $m_{H^+} < $ 300 GeV, and from
$\tilde\chi^+$ and $\tilde\chi^0$ for $m_{H^+} \geq $ 300 GeV.
 The dependence on different values of $\tan\beta$ is examined.

The approximate number of $H^\pm$'s needed to measure
$\delta_{Wh^0}$ is $N_{H^\pm}\geq 1/[\delta_{Wh^0}^2\,Br (H^+\to
W^\pm h^0)]$, which, for $\delta_{Wh^0} \simeq 10^{-2}$ and a
branching ration  $\simeq 10\%$, implies $N_{H^\pm}\geq 10^5$.

 Charged $H^\pm$ will be produced at the Tevatron in FermiLab
 if $m_{H^+}\leq$ 300 GeV, and at LHC in CERN
 if $m_{H^+}\leq$ 1000 GeV. As the cross sections for $H^\pm$ production at $pp$ and $p\bar p$
collisions decreases strongly for low $\tan\beta$, the required
number $N_{H^\pm}$ is too large for the planned luminosities at
the hadron colliders \cite{Belyaev}. For example, at LHC with
integrated luminosity per year $L$ = 100 $fb^{-1}$, at $m_{H^+}$ =
500 GeV and $\tan\beta = 10$, with an efficiency for the signal
$\epsilon = 2,6\%$, using the results of \cite{Belyaev}, for the
ratio of the signal ($S$) and the background $(B)$ events, we
obtain $S/B \simeq 65/3770$.

 More promising are the  linear $e^+e^-$ colliders.
 In this case the charged Higgs will be copiously produced,
 the main production mechanism being
 $e^+e^- \to H^+H^-$. Thus, the only parameter for
 the production cross section, at tree level, is the Higgs boson mass $m_{H^+}$.
For a collider at $\sqrt s$ = 800 GeV with luminosity $L$ = 500
$fb^{-1}$, the cross section is $\sim $ 29 $fb$ for $m_{H^+}$ =
200 GeV, and $\sim $ 12 $fb$ for $m_{H^+}$ = 300 GeV, which
corresponds to 1.5 $\times$ $10^4$ and  6 $\times$ $10^3$
$H^+H^-$-pairs, respectively \cite{Kiiskinen}. For the CLIC
collider, at $\sqrt s$ = 3 TeV,  the cross section is 3 $fb$ for
$m_{H^+} = 400$ GeV which, for $L$ = 800 $fb^{-1}$, corresponds to
2.4 $\times$ $10^3$ charged Higgs pairs.
 This implies
 that at the NLC  higher luminosities will be needed
 for such an asymmetry to be measured.

\section*{Acknowledgements}

We are thankful to Helmut Eberl for his help with the branching
ratios of the charged Higgs. This work is supported by the
Bulgarian National Science Foundation,
 Grant Ph-1010/00.


\begin{appendix}
\section{ Lagrangian with complex couplings}

The mass matrices and their diagonalization matrices are defined
in the Appendicies A of \cite{1} -- for charginos and neutralinos
  and of \cite{2} -- for staus and
sneutrinos. Here we give only the pieces of the interaction
Lagrangian we use.\\

{\bf 1. Lagrangian with neutral Higgses}
\be
{\cal L}_{h^0\,\tilde{\tau}_m^* \tilde{\tau}_n}&=& g\sum_{m,n=1,2}
c_{mn}^{\tilde\tau} \tilde{\tau}_m^* \tilde{\tau}_n\,h^0\,,
\qquad c_{mn}^{\tilde\tau}=c_{nm}^{\tilde\tau *}\\
{\cal L}_{h^0\,\widetilde{\nu}\widetilde{\nu}}&=&
g\,{m_z\over 2\cos \theta_W}\sin(\alpha+\beta)\,\widetilde{\nu}_L^*\widetilde{\nu}_L\,h^0\\
{\cal L}_{h^0\tilde\chi^0_l\tilde\chi^0_k} &=& g\,\bar{\tilde{\chi}}^0_l
\left(A_{lk}^*\,P_L + A_{lk}\,P_R\right) \tilde\chi^0_k\,h^0\\
{\cal L}_{h^0\tilde\chi^+_i\tilde\chi^+_j} &=&
g\,\bar{\tilde{\chi}}^+_i \left(\widetilde{A}_{ij}^*\,P_L +
\widetilde{A}_{ji}\,P_R\right) \tilde\chi^+_j\,h^0\\
{\cal L}_{h^0WW} &=& g m_W\sin(\beta-\alpha)W_{\mu}^+W^{-\mu}h^0
 \ee
where
\be c_{mn}^{\widetilde{\tau}}=c_{LL}^{\widetilde{\tau}}{\cal
R}_{Lm}^{\widetilde{\tau}*} {\cal
R}_{Ln}^{\widetilde{\tau}}+c_{RR}^{\widetilde{\tau}} {\cal
R}_{Rm}^{\widetilde{\tau}*}{\cal R}_{Rn}^{\widetilde{\tau}}+
c_{RL}^{\widetilde{\tau}}{\cal R}_{Rm}^{\widetilde{\tau}*}{\cal
R}_{Ln}^{\widetilde{\tau}}+ c_{LR}^{\widetilde{\tau}}{\cal
R}_{Lm}^{\widetilde{\tau}*}{\cal R}_{Rn}^{\widetilde{\tau}} \nee
\be c_{LL}^{\widetilde{\tau}}&=&{m_z\over \cos \theta_W}(-{1\over
2}+{\sin^2 \theta_W}) \sin(\alpha+\beta)+{m_{\tau}^2\over
m_W}{\sin \alpha\over \cos \beta},\nn
c_{RR}^{\widetilde{\tau}}&=&-{m_z\over \cos \theta_W}\sin^2
\theta_W \sin(\alpha+\beta) +{m_{\tau}^2\over m_W}{\sin
\alpha\over \cos \beta},\nn
c_{LR}^{\widetilde{\tau}}&=&{m_{\tau}\over 2m_W\cos\beta}(\mu
\cos\alpha +A_{\tau}^*\sin\alpha) \nn
c_{RL}^{\widetilde{\tau}}&=&{m_{\tau}\over {2m_W\cos \beta
}}(\mu^*\cos\alpha +A_\tau\sin\alpha )=c_{LR}^{\widetilde{\tau}
*} . \nee
\be
A_{lk}&=&{1\over 2}(\sin\alpha \,Q_{lk}^{''} + \cos\alpha\, S_{lk}^{''}),\nn
Q_{lk}^{''}&=&{1\over 2}[N_{l3}(N_{k2}-N_{k1}\tan \theta_W)+(l\longleftrightarrow k)],\nn
S_{lk}^{''}&=&{1\over 2}[N_{l4}(N_{k2}-N_{k1}\tan\theta_W)+(l\longleftrightarrow k)],
\nee
\be \widetilde{A}_{ij}&=& \sin\alpha\, Q_{ij}- \cos\alpha \,
S_{ij},\nn Q_{ij} &=&{1\over \sqrt{2}}U_{i2}V_{j1},\qquad S_{ij}
={1\over \sqrt{2}}U_{i1}V_{j2}. \nee
\nn
{\bf 2. Lagrangian with charged  Higgses}\footnote{The
correspondence with the notation in refs.\cite{1} and \cite{2} is
$gf^{L,R} = F^{L,R}$, $ g/(\sqrt 2 m_W) g_4^{\tilde{\tau}} =
G_4^{\tilde{\tau}}$, $ g/(\sqrt 2 m_W) g_4^{\tilde{t}} =
G_4^{\tilde{t}}$}
\be
{\cal L}_{h^0H^+W^-} &=&-\frac{ig}{2} \cos (\alpha -\beta ) \left\{(H^-
\plr_{\,\alpha} \,h^0)\,W^{+,\alpha} -
(H^+\, \plr_{\,\alpha} \,h^0)\,W^{-,\alpha}\right\}\\
{\cal L}_{H^+\tilde\chi^+_j\tilde\chi^0_k } &=&
g\left\{H^-\bar{\tilde\chi}^{\,0}_k\,
 (\, f_{kj}^{R*}P_L + f_{kj}^{L*}\,P_R\,)\,\tilde\chi^+_j
+H^+\bar{\tilde\chi}^+_j
 \,(\, f_{jk}^{R}P_R + f_{jk}^{L}\,P_L\,)\,\tilde\chi^{\,0}_k\right\}\\
 {\cal L}_{H^+\tilde{\tau}_n\tilde{\nu}}&=&
 \frac{g}{\sqrt 2\,m_W}\left[ (g_4^{\widetilde{\tau}})_nH^+ \tilde \nu_\tau^*\tilde\tau_n+
 (g_4^{\widetilde{\tau}})^*_n
 H^-\tilde\tau_n^*\tilde\nu_\tau\right]
\ee
where
 \be
  \qquad f_{kj}^L&=&-\sin\beta\left[N_{k3}^*U_{j1}^* -
\frac{1}{\sqrt 2}(N_{k2}^* +N_{k1}^*
\tan\theta_W)U_{j2}^*\right]\nn \qquad
f_{kj}^R&=&-\cos\beta\left[N_{k4}V_{j1} + \frac{1}{\sqrt
2}(N_{k2}  +N_{k1}  \tan\theta_W)V_{j2} \right]   \nn
(g_4^{\widetilde{\tau}})_n &=& a_{LL}^{\widetilde{\tau}}{\cal
R}_{Ln}^{\tilde\tau} +
 a_{LR}^{\widetilde{\tau}}
{\cal R}_{Rn}^{\tilde\tau}\,, \nn
 a_{LL}^{\widetilde{\tau}} &=&
m_\tau^2\tan\beta - m_W^2\sin 2\beta   \nn
a_{LR}^{\widetilde{\tau}} &=& m_\tau (A_\tau^*\tan\beta +\mu )
\nee

  {\bf 3. Lagrangian with $W^\pm $}
 \be
  {\cal L}_{W^+\tilde
\chi^+_j  \tilde \chi^0_k } &=& g \left\{\bar{\tilde\chi}^0_k\,
\gamma^\alpha \,(O_{kj}^L P_L + O_{kj}^R P_R)\, \tilde \chi^+_j
\,W_{\alpha}^- +\bar{\tilde\chi}^+_j\, \gamma^\alpha
\,(O_{jk}^{L*} P_L +
 O_{jk}^{R* }P_R)\,
\tilde \chi^0_k \,W_{\alpha}^+\right\}\nn\\
{\cal L}_{W\tilde \nu \tilde \tau_n } &=&\frac{-i g}{\sqrt 2}
\left\{{\cal R}_{Lm}^{\tilde \tau *}\,W^-_\alpha \left(\tilde
\tau_m^{*} \plr^{\,\alpha} \,\tilde \nu \right) + {\cal
R}_{Lm}^{\tilde \tau}\,W^+_\alpha \left(\tilde \nu^{*}
\plr^{\,\alpha}\, \tilde \tau_m\right)\right\}
\ee
\\ Here
 \be\quad O_{kj}^L &=& -\,\frac{1}{\sqrt
2}\, N_{k4}\,V_{j2}^* + N_{k2}\,V_{j1}^*,\qquad O_{kj}^R =
\frac{1}{\sqrt 2}\, N_{k3}^*\,U_{j2} + N_{k2}^*\,U_{j1} \nee

\section{ Absorptive parts of the Integrals}

If we use the notation:
\be
{\cal D}^0=q^2-m_0^2,\qquad {\cal D}^j=(q+p_j)^2-m_j^2,
\nee
then the Passarino-Veltman two- and three-point
functions \cite{PV}, when the loop integrals are given in
4-dimentions,  are:
\be B_0(p_1^2, m_0^2, m_1^2)&=&{1\over i\pi^2}\int
d^4q{1\over{\cal
D}^0{\cal D}^1 },\\
C_0(p_1^2, (p_1-p_2)^2, p_2^2, m_0^2, m_1^2, m_2^2) &=&{1\over
i\pi^2}\int d^4q{1\over {\cal D}^0{\cal D}^1{\cal D}^2},\\
C_{\mu}(p_1^2, (p_1-p_2)^2, p_2^2, m_0^2, m_1^2, m_2^2) &=&{1\over
i\pi^2}\int d^4q{q_{\mu}\over {\cal D}^0{\cal D}^1{\cal
D}^2}=p_{1\mu}C_1+p_{2\mu}C_2
\ee
For our diagrams in the considered $H^+$ decay we  have:
 \be
p_1=p_{H^+} \Rightarrow p_1^2 =m_{H^+}^2,\qquad p_2=p_h
\Rightarrow p_2^2 =m_h^2,\qquad (p_1-p_2)^2 =m_W^2.
 \nee
 Then for the the absorptive  parts of the
integrals, when the particles with masses $m_0$ and $m_1$ are put on mass shell
we obtain:
\be \Im m\, B_0( m_{H^+}^2,m_0^2, m_1^2)&=&\frac{2\pi \vert
\overrightarrow{k}\vert}{m_{H^+}}\\
\Im m \,C_0(m_{H^+}^2, m_W^2, m_h^2, m_0^2, m_1^2,m_2^2)&=&
 \frac{-\,\pi}{2\,m_{H^+} \vert \overrightarrow{p_h}\vert}\, \ln \vert\frac{a+b}{a-b}\vert \\
\Im m \,C_1( m_{H^+}^2 , m_W^2,m_h^2,m_0^2,
m_1^2,m_2^2)&=&\frac{m_h^2\, A
-(p_{H^+}p_h)\, B}{\Delta}\\
\Im m \,C_2(m_{H^+}^2, m_W^2, m_h^2, m_0^2,
m_1^2,m_2^2)&=&\frac{-(p_{H^+}p_h)\,A + m_{H^+}^2\, B}{\Delta}.
\ee
Here
 \be
 \Delta &=& m_{H^+}^2m_h^2 -(p_{H^+}p_h)^2,\qquad  (p_{H^+}p_h)=
\frac{m_{H^+}^2 +m_h^2 -m_W^2}{2}\nn \vert \overrightarrow{k}\vert
&=&\frac{\lambda^{1/2}(m_{H^+}^2,m_0^2,m_1^2)}{2m_{H^+}},\qquad
\vert \overrightarrow{p_h}\vert
=\frac{\lambda^{1/2}(m_{H^+}^2,m_h^2,m_W^2)}{2m_{H^+}},\nn \lambda
(x,y,z)&=& x^2+y^2+z^2 -2xy-2xz-2yz\nn A
&=&\frac{-\,\pi\,k^0}{2\, \vert \overrightarrow{p_h}\vert}\, \ln
\vert\frac{a+b}{a-b}\vert ,\qquad B =\frac{-\,\pi
}{2\,m_{H^+}\,\vert \overrightarrow{p_h}\vert}\,
 \left\{\left[k^0\,p_h^0 -\frac{a}{2}\right]
  \ln \vert\frac{a+b}{a-b}\vert
  -2\, \vert \overrightarrow{k}\vert \,\vert \overrightarrow{p_h}\vert\right\}\nn
a&=& m_h^2+m_0^2 -m_2^2 + 2\,p_h^0k^0,\qquad b= -2 \vert
\overrightarrow{k}\vert\,\vert \overrightarrow{p_h}\vert\,,\nn
k^0&=&\frac{m_1^2-m_0^2 -m_{H^+}^2}{2m_{H^+}},\qquad p_h^0=
\frac{m_{H^+}^2-m_W^2 + m_h^2}{2m_{H^+}}. \nee

\end{appendix}
\newpage

\end{document}